# Gapped commensurate antiferromagnetic response in a strongly underdoped model cuprate superconductor

Z. W. Anderson[1,*], Y. Tang[1], V. Nagarajan[1], M. K. Chan[1,†], C. J. Dorow[1], G. Yu[1], D. L. Abernathy[2], A. D. Christianson[3], L. Mangin-Thro[4], P. Steffens[4], T. Sterling[5], D. Reznik[5], D. Bounoua[6], Y. Sidis[6], P. Bourges[6], and M. Greven[1,*]

[1] *School of Physics and Astronomy, University of Minnesota, Minneapolis, MN 55455, USA*
[2] *Neutron Scattering Division, Oak Ridge National Laboratory, Oak Ridge, TN 37831, USA*
[3] *Materials Science and Technology Division, Oak Ridge National Laboratory, Oak Ridge, TN 37831, USA*
[4] *Institut Laue-Langevin, 71 avenue des martyrs, Grenoble 38000, France*
[5] *Department of Physics, University of Colorado Boulder, Boulder, Colorado 80309, USA*
[6] *Laboratoire Léon Brillouin, CEA-CNRS, Université Paris-Saclay CEA-Saclay, Gif sur Yvette 91191, France*
[†] *Present address: Pulsed Field Facility, National High Magnetic Field Laboratory, Los Alamos National Laboratory, Los Alamos, New Mexico 87545, USA*
* and01942@umn.edu, greven@umn.edu

**It is a distinct possibility that spin fluctuations are the pairing interactions in a wide range of unconventional superconductors. In the case of the high-transition-temperature (high-$T_c$) cuprates, in which superconductivity emerges upon doping an antiferromagnetic Mott-insulating state, spin correlations might furthermore drive unusual pseudogap phenomena. Here we use polarized and unpolarized neutron scattering to study the simple tetragonal cuprate HgBa$_2$CuO$_{4+\delta}$ at very low doping ($T_c \approx 55$ K, hole concentration $p \approx 0.064$). In stark contrast to prior results for other underdoped cuprates, we find no evidence of incommensurate spin-density-wave, charge-spin stripe, or q = 0 magnetic order. Instead, the antiferromagnetic response in both the superconducting and pseudogap states is gapped below $\Delta_{AF} \approx 6$ meV, commensurate over a wide energy range, and disperses above about 55 meV. Given the documented model nature of HgBa$_2$CuO$_{4+\delta}$, which exhibits high structural symmetry and minimal point disorder effects, we conclude that the observed behavior signifies the unmasked response of the quintessential CuO$_2$ planes near the Mott-insulating state. These results for HgBa$_2$CuO$_{4+\delta}$ can therefore be expected to serve as a benchmark for a refined theoretical understanding of the cuprates.**

The cuprate superconductors share a quintessential structural and electronic unit: the copper-oxygen plane. The parent insulating state is characterized by a large Cu $3d$ - O $2p$ charge-transfer gap (~ 1-2 eV) [1], a spin-1/2 Cu moment [2], and quasi two-dimensional (2D) antiferromagnetic (AF) spin correlations with a large superexchange energy (~ 130 meV) [3,4]. Superconductivity emerges upon doping the planes with about $p \sim 0.05$ holes per planar Cu. Below optimal doping ($p \sim 0.16$), where $T_c$ is largest, the superconducting (SC) phase is preceded by the pseudogap (PG) region at higher temperature (Fig. 1); the latter has been the subject of much research activity and found to exhibit myriad ordering phenomena (translational-symmetry-preserving $q = 0$ magnetism, charge-density-wave, charge-spin stripe, electron-nematic) [5]. Since spin correlations might drive both superconductivity [6,7] and some of the observed PG phenomena [8-10], it is imperative to determine the magnetic response of the pristine $CuO_2$ planes. Indeed, the past four decades have seen tremendous efforts toward this end. Regardless of the nature of the pairing mechanism, a successful theory of the cuprates ought to capture the spin dynamics of these quantum materials.

Neutron scattering is a powerful probe of the full momentum- and energy-dependent magnetic response of a material, i.e., the scattering cross section is proportional to the imaginary part of the dynamic magnetic susceptibility, $\chi''(\mathbf{Q},\omega)$, where $\mathbf{Q}$ and $\omega$ are the momentum and energy transfer to the scattering system, respectively. Due to limitations in the attainable flux even at the best neutron sources, a measurement of the spin dynamics typically requires sizable crystals, especially in the case of low-dimensional systems and in the extreme quantum limit of spin-1/2, due to sizable quantum fluctuations. For the cuprates, this has largely constrained studies of the magnetic response to two systems for which such crystals have been available [11-14]: $La_{2-x}Sr_xCuO_4$ (LSCO) and related compounds, as well as $YBa_2Cu_3O_{6+\delta}$ (YBCO). Key features revealed by these studies are: (i) in YBCO, a normal-state excitation gap ($\Delta_{AF}$) at the antiferromagnetic wave vector that increases with increasing doping in the SC doping range; (ii) an hourglass-shaped normal-state dispersion that is less pronounced in YBCO than in LSCO; (iii) in the case of YBCO, a resonance upon cooling into the SC state; (iv) near $p \sim 1/8$, LSCO and the closely related $(La,Nd)_{2-x}(Sr,Ba)_xCuO_4$ compounds exhibit a pronounced instability toward stripe order, with incommensurate magnetic correlations, that is stabilized in a low-symmetry structural phase. Incidentally, below $p \sim 0.085$, YBCO exhibits quasi-static incommensurate spin-density-wave (SDW) correlations [15]. Whereas the high-energy response (the upper part of the hourglass) in both compounds resembles the spin-wave dispersion seen in the insulating parent state, the considerable differences observed below about 50 meV have raised questions regarding the nature of the underlying magnetic response of the $CuO_2$ planes in the absence of compound-specific idiosyncrasies. This issue has not been resolved through more limited neutron scattering studies of several other cuprates [12,13] - $Bi_2Sr_2CuO_{6+x}$ (Bi2201), $Bi_2Sr_2CaCu_2O_{8+x}$ (Bi2212) and $Tl_2Ba_2CuO_{6+\delta}$ (Tl2201). Note that 50 meV is an important energy scale, as it is comparable to both the PG and SC gap scales near optimal doping [16].

There are good reasons to think that $HgBa_2CuO_{4+\delta}$ (Hg1201) is a model system [17-19], as it features the highest optimal $T_c$ (nearly 100 K) of all single-$CuO_2$-layer cuprates (such as LSCO, Bi2201, Tl2201), higher also than the corresponding values for double-layer YBCO and Bi2212 [20]. Moreover, unlike YBCO and LSCO, which exhibit low orthorhombic structural symmetry (in the entire SC doping range of YBCO, and up to $p \sim 0.22$ in the case of LSCO), Hg1201 maintains high tetragonal symmetry [20]. Finally, point disorder effects are rather prominent in LSCO and cause an insulator-like low-temperature resistivity upturn below optimal doping. Such effects have not been observed for Hg1201 in samples with doping levels as low as $p \sim 0.055$ ($T_c \sim 45$ K) [22,23], and although less prominent in YBCO, they are present in the doping range where SDW correlations are observed (below $p \sim 0.085$) [24]. Notably, in YBCO, both the resistivity upturn [25] and the incommensurate magnetic order [26] are dramatically enhanced (up to at least $p = 0.12$) with intentionally introduced point disorder. The degree to which the underlying electronic properties of the $CuO_2$ planes can be masked by low structural symmetry and/or point disorder effects (e.g., through the stabilization of secondary electronic phases) is exemplified by a study [19] that uncovered that: Hg1201 obeys conventional Kohler scaling of the magnetoresistance deep in the PG state, with a Fermi-liquid $T^2$ scattering rate; in the case of YBCO, this simple metallic behavior is only seen in de-twinned samples, for charge-current flow perpendicular to the Cu-O chains characteristic of this particular compound; in LSCO, Bi2201 and Bi2212, this simple underlying scaling behavior is masked even more strongly. In this context, it is important to note that theoretical treatments of the cuprates generally avoid consideration of the added complexity due to non-universal disorder effects and deviations from tetragonal symmetry.

In recent years, sizable samples of Hg1201 have become available. Initial neutron scattering work revealed a gapped AF response near optimal doping ($p \sim 0.13$, $T_c = 88$ K; sample denoted Hg1201-UD88) similar to YBCO, with a resonance and an hourglass-shaped dispersion in the SC state [27]. In stark contrast, a second study at somewhat lower doping ($p \sim 0.09$, $T_c = 71$ K; Hg1201-UD71) revealed an unusual, gapped wineglass-shaped response that is hardly altered upon cooling into the SC state [28]. Since CDW correlations are particularly robust at the doping level of the latter study (Fig. 1), it has remained unclear if the observed behavior is a mere peculiarity associated with the prominence of the CDW phenomenon. It is therefore highly desirable to determine the magnetic response of Hg1201 at lower doping, where CDW order is weak [29].

Here, we present a detailed neutron scattering study of a strongly underdoped Hg1201 sample ($p \approx 0.064$ and $T_c \approx 55$ K; Hg1201-UD55) and uncover the underlying magnetic response of the $CuO_2$ planes below $\sim 140$ meV near the parent insulating state. As in the prior measurements of the AF excitations in Hg1201 at higher doping, we primarily report time-of-flight results [27,28]. In order to confirm the magnetic nature of the response, we also present complementary triple-axis data with neutron polarization analysis. We find dynamic, wine-glass shaped AF correlations with little change across $T_c$, closely similar to the prior result for Hg1201-UD71, but with a much smaller excitation gap of $\Delta_{AF} \approx 6$ meV. Furthermore, we observe no signs of stripe/SDW and $q = 0$ magnetic order. Motivated by recent neutron scattering results for LSCO [30], we carry out density-functional theory (DFT) phonon calculations to address the possibility that the spin excitations in Hg1201 are coupled to the lattice dynamics at select energies. We find no evidence of significant intrinsic lattice effects.

## Results

Figure 2a shows contour plots of $\chi''(\mathbf{Q}, \omega)$ at select energy transfers $\omega$ and temperatures $T = 5$ K, 70 K and 410 K, obtained with incident neutron energies $E_i = 70$ and 200 meV using the same data analysis method as described previously [27,28] and in the Supplementary Information. We quote the three-dimensional scattering wave-vector $\mathbf{Q} = H\mathbf{a}^* + K\mathbf{b}^* + L\mathbf{c}^* \equiv (H\ K\ L)$ in reciprocal lattice units, where $a^* = b^* = 1.62$ Å$^{-1}$ and $c^* = 0.66$ Å$^{-1}$ are the room-temperature values. Given the lamellar nature of the cuprates, we are interested in the 2D magnetic response. Data such as those in Fig. 2a are therefore fit to the heuristic gaussian function $\chi''(\mathbf{Q}) = \chi_0'' \exp\{-4ln2\ R/(2\kappa)^2\}$, where $R = |[(H - 1/2)^2 + (K - 1/2)^2]^{1/2} - \delta|^2$, $2\kappa$ is the intrinsic full-with-at-half-maximum (FWHM) momentum width, and $\delta$ parameterizes the incommensurability relative to the 2D AF wave vector $q_{AF} = (1/2\ 1/2)$. Corresponding constant-$\omega$ momentum trajectories, averaged over {100} and {010} with lateral momentum-bin range $q_\perp = [0.45, 0.55]$ r.l.u., are shown in Fig. 2b. The overall response at $T = 5$ K and 70 K, extracted from the fits up to ~135 meV, is shown in Fig. 2c. $\chi''(\mathbf{Q}, \omega)$ changes from dispersing to commensurate below $\omega_c \approx 55$ meV, and becomes immeasurably small below $\Delta_{AF} \approx 6$ meV. At 410 K, the response is considerably weaker, yet consistent with these observations.

Figures 3a,b show, respectively, the energy dependence of the amplitude $\chi_0''(\omega)$ and the change $\Delta\chi_0''(\omega)$ across $T_c$ and between 5 K and 410 K. Figures 3c,d show the corresponding results for the momentum-integrated local susceptibility, defined as $\chi_{loc}''(\omega) = \int \chi''(\mathbf{q}, \omega) d^2q / \int d^2q$, and its change $\chi_{loc}''(\omega)$. There is no notable difference between the response at 5 K and 70 K, except for a subtle low-temperature enhancement at $\omega \sim 30$ meV, where $\chi_0''(\omega)$ is largest. We notice a spike at ~ 30 meV in $\chi_{loc}''(\omega)$ in the 5 K and 70 K data, and broad maxima at ~ 30 meV and ~ 50 meV in the 410 K data. Such enhancements were previously observed and ascribed to additional phonon scattering that is not properly removed in the processed data (see Supplementary Information) [28]. The results in Figs. 2 and 3 reveal that the magnetic response is gapped already in the normal state, with no significant change across $T_c$. At 5 K and 70 K, $\chi_0''(\omega)$ and $\chi_{loc}''(\omega)$ are peaked at ~ 35 meV and ~ 50 meV, respectively. Above $\omega \sim 100$ meV, the response is relatively weak and rather similar at all three temperatures.

In order to obtain data with even better energy resolution and signal-to-noise ratios in the energy range in which the response is commensurate, we carried out measurements with $E_i = 50$ meV. Figure 4a shows $\chi_0''(\omega)$ below 45 meV at seven temperatures in the 10 K to 450 K range. We find again that the magnetic response extrapolates to zero at $\Delta_{AF} \approx 6$ meV, both in the normal state and deep in the SC state. As already seen in Fig. 3a for 5 K and 70 K, $\chi_0''(\omega)$ exhibits a smooth, non-monotonic energy dependence at all temperatures. The $E_i = 50$ meV results are consistent with those obtained with higher incident energies (see Fig. 4a inset) within about 15%. Figure 4c shows the same data as a function of temperature. At high energy transfers, $\chi_0''$ smoothly increases with decreasing temperature and begins to plateau near $T_c$. For $\omega \leq 12$ meV, $\chi_0''$ eventually decreases at low temperatures, consistent with the opening of a gap.

Figure 4b shows the energy dependence of the response at $\mathbf{q}_{AF}$ obtained from triple-axis momentum and energy scans with longitudinal polarization analysis, obtained at 5 K and 70 K. These data agree well with the time-of-flight result, and hence confirm the predominant magnetic nature of the scattering and the presence of a gap.

We have tested the possibility of quasi-elastic magnetic scattering, which is seen in other cuprates. Figure 5a shows triple-axis [0 1 0] momentum scans across $\mathbf{q}_{AF}$ at nominally zero energy transfer (1.2 meV FWHM energy resolution). The SC (5 K) and normal (80 K) state data are featureless and statistically indistinguishable. Importantly, unlike for YBCO and LSCO [11-13], where strong incommensurate quasi-elastic SDW peaks have been observed at similar doping levels, we observe no such scattering in Hg1201-UD55, consistent with our observation of a gapped response centered at $\mathbf{q}_{AF}$. Using the known cross section of the (0 0 4) Bragg peak (which is 4.2 barn) for comparison, we estimate the upper bound of a possible elastic AF signal to be 0.4 mbarn.

Finally, we also investigated the existence of a $q = 0$ magnetic response, as previously observed at higher doping [31-33]. Figure 5b shows the results of polarized neutron diffraction measurements of the magnetic scattering at the (1 0 0) Bragg peak in Hg1201-UD55 compared to earlier results [33] for Hg1201-UD71 (moderately-underdoped) and Hg1201-OP95 (optimally-doped). All data were obtained with the D7 spectrometer at ILL. For Hg1201-UD71, the magnetic scattering (extracted from XYZ-longitudinal polarization analysis - see the Supplementary Information) exhibits an order-parameter-like temperature dependence with an onset below $T_0 \sim 370$ K and a maximum intensity of ~ 13 mbarn at 80 K. Conversely, the highly underdoped and optimally-doped samples show no such signature of $q = 0$ magnetism within the detection limit of the instrument. However, we note that a nonzero $q = 0$ magnetic response at both the (1 0 0) and (1 0 1) reflections with a magnitude of ~ 1.7 mbarn at 100 K was previously observed for Hg1201-OP95 using a triple-axis neutron spectrometer (4F1 at LLB) [33]. Due to the thermal drift of the intensities collected in the different measurement channels (see Supplementary Information and Ref. [33]) and the inability to determine an accurate baseline for the magnetic scattering, the weak $q = 0$ signal falls below the threshold of detection on D7. One can thus give an upper bound of 1.7 mbarn for the $q = 0$ magnetism in Hg1201-UD55, as indicated by the shaded area in Fig. 5b. Interestingly, the vanishing of the $q = 0$ magnetic signal at low doping in Hg1201 is consistent with the significant decrease observed in underdoped $YBa_2Cu_3O_{6.45}$ ($p = 0.08$) [34].

## Discussion

At the low doping levels of the present study, the cuprates YBCO, LSCO, and Bi2201 exhibit an ungapped, hourglass-shaped magnetic response in both the normal and SC states [11,13,15,35,36]. In contrast, our results for Hg1201-UD55 resemble the wineglass-shaped response previously observed for Hg1201-UD71, albeit with a considerably smaller gap (~ 6 meV vs ~ 27 meV). These findings are crucial, for two reasons. First, given that CDW order in Hg1201 is nearly absent at the doping level of the present study [29], the prior result for Hg1201-UD71 does not appear to be affected by CDW correlations.

Second, the gapped, wineglass-shaped response must be the intrinsic behavior of the pristine, quintessential $CuO_2$ planes. Unlike other hole-doped cuprates that have been investigated in this context, Hg1201 exhibits high tetragonal symmetry and minimal disorder effects, as evident from transport measurements, e.g., the observations of Kohler scaling, negligible residual resistivity, and quantum oscillations in the underdoped part of the phase diagram [17-19,37]. Incommensurate SDW and stripe correlations at low and intermediate hole concentrations [5,15,35] therefore are not universal features of the cuprates, but rather the result of material-specific interactions. While the high-energy PG energy scale is ~ 400 meV at the doping level of the present study [38,39], the SC gap is expected to be well below the characteristic scale $\omega_c \approx 55$ meV above which the response of Hg1201-UD55 becomes incommensurate [40]. The present results, which extend to $\omega \sim 140$ meV, thus cover an important energy range and provide pivotal information for tests of SC pairing scenarios, such as spin-fluctuation-driven mechanisms [6].

At somewhat higher doping levels than in the present work, Hg1201 and several other cuprates exhibit a resonance-like enhancement at $\mathbf{q}_{AF}$ (and characteristic energy $\omega_r$) in the SC state, with a response that is gapped and hourglass- rather than wineglass-shaped (Fig. 1). This can be understood within an itinerant picture, in which the resonance is part of a spin exciton, i.e., a dispersive spin-triplet collective mode bound below the threshold of the Stoner continuum [27]. Unlike for Hg1201-UD88, which exhibits a resonance at $\omega_r \approx 59$ meV [27], no resonance was observed in Hg1201-UD71 [28]. In the present work, we find an increase of the magnetic response in the 25-35 meV range in the SC state, yet it is unclear if this small, gradual increase is to be thought of as a resonance. From the data in Figs. 3 and 4, we estimate this spectral weight enhancement to be $W_r = \int d\omega \, \Delta\chi''_{loc} = 0.14 \pm 0.07 \, \mu_B^2 \text{ f.u.}^{-1}$. Additional evidence for a small resonance-like effect from triple-axis scattering data is shown in Fig. S9 in the Supplementary Information. Although the value of the SC gap at the doping level of Hg1201-UD55 is unknown, it is likely not much larger than $\omega_r \approx 30$ meV [40], so that $W_r$ is indeed expected to be small [27]. Moreover, the ratio $\omega_r/T_c \sim 6.4$ is comparable to that for other cuprates [41]. We caution though that several phonon modes cross $q_{AF}$ at about 30 meV, as shown in detail in Fig. 6 and also discussed in [27,28], which complicates the interpretation of subtle changes in the magnetic response across $T_c$.

Our polarized triple-axis neutron scattering data extend up to about 40 meV and confirm that the low-energy commensurate response is predominantly magnetic in nature (Fig. 4b). However, in the time-of-flight data, both the peak and local susceptibilities at 410 K exhibit local maxima at energies (about 30 meV and 50 meV) that correspond to phonon crossings (Fig. 6), and at low temperatures, $\chi_0''$ is strongest at ~ 35 meV (Figs. 2-4). This points to the possibility of nontrivial electron-lattice coupling effects. A recent neutron scattering study of LSCO revealed an enhancement of $\chi''(\omega)$ in the 16-19 meV range, where the (incommensurate) spin excitations are intersected by optic phonons [30]. It was found that this enhancement is present only in SC samples and that, similar to the transition temperature, its magnitude follows a dome-shaped doping dependence. This effect was interpreted as an interplay among spin, charge, and lattice degrees of freedom, with resultant composite excitations: optic phonons stabilize fluctuating, short-range stripe correlations, and thereby enhance the magnetic correlations. Although Hg1201 does not display a tendency toward stripe order, similar electron-lattice coupling effects may be present. Since we lack data for non-SC samples for comparison, we are unable to carry out the same analysis as Ref. [30]. Instead, we have performed a DFT calculation of the phonon spectrum of stoichiometric, undoped Hg1201. The observed good agreement between experiment and theory (Fig. 6) allows us to identify the approximate energy ranges that are unaffected by phonon crossings at $q_{AF}$. For example, only the $\omega = 30$ meV data in Fig. 2a,b are affected, due to the relatively strong dispersion and $Q$-dependence of a phonon mode crossing $q_{AF}$ at this energy; this contribution is not fully removed during our data analysis. Similarly, most of the energy transfers for which data are shown in Fig. 4b do not coincide with calculated phonon crossings. From this, we can conclude that, indeed, lattice effects play an overall minor role.

Comparing the present results with those for Hg1201-UD71 and Hg1201-UD88, we find that the low-temperature maximum of the local susceptibility, $\chi_{loc}''(\omega) \sim 4\text{-}5\ \mu_B^2\,\text{eV}^{-1}\text{f.u.}^{-1}$, is approximately independent of doping. Similarly, the strength of the total, energy-integrated response (up to 100 meV) is comparable in Hg1201-UD55 and Hg1201-UD71 [28], and about 0.25-0.3 $\mu_B^2$ f.u.$^{-1}$; extrapolating the available data for Hg1201-UD88 [27] to 100 meV yields 0.15-0.2 $\mu_B^2$ f.u.$^{-1}$. Despite the differences in the structure of the response (wineglass vs. hourglass), this compares rather well with the corresponding strength of the magnetic response of underdoped LSCO (~ 0.15 $\mu_B^2$ f.u.$^{-1}$ for $p = 0.085$) [42].

Similar to Hg1201-UD71 [28], Hg1201-UD88 [27], and underdoped YBCO [43-45], the AF response in Hg1201-UD55 starts to increase below the PG temperature (Figs. 1 and 4c). This observation can be understood within the framework of a recent phenomenological model for the cuprates [38,39], which combines insights from transport measurements with evidence that these complex oxides are inherently inhomogeneous and exhibit local two-component electronic behavior: Mott-localization of one hole per Cu occurs gradually with decreasing doping and/or temperature, in a highly inhomogeneous manner, such that below $T^*$ (or, more precisely, the somewhat lower temperature $T^{**}$), the carrier density has fully crossed over from $1 + p$ to the nominal value $p$. In this picture of the phase diagram, the formation of local magnetic moments, loop currents, CDW,

SDW and stripe order are all emergent phenomena, and significant (dynamic) AF correlations are expected to develop below $T^*$. While this comprehensive model is phenomenological in nature, the present results pave the way for a microscopic theoretical understanding of the magnetic degrees of freedom of the doped $CuO_2$ planes near the Mott-insulating parent state.

## Methods

### Neutron scattering experiments

The neutron scattering measurements were performed on a large co-aligned sample (mass of ~ 2 g) that consists of approximately 30 single crystals. Neutron diffraction on the full sample shows a FWHM mosaic of 1.7°. Crystals were grown by a flux method [46] and then annealed to the desired doping level at an elevated temperature in vacuum [37]. For each crystal, $T_c$ was determined from a magnetic susceptibility measurement using a Quantum Design, Inc. Magnetic Property Measurement System. The average of these measurements was used to characterize the full sample, resulting in a mean value of $T_c = 55$ K and a FWHM transition width of $\Delta T_c = 8$ K.

The measurements of the AF response were carried out on the time-of-flight spectrometer ARCS at Oak Ridge National Laboratory (ORNL), Oak Ridge (USA), and on two triple-axis spectrometers: IN20 at the Institut Laue-Langevin (ILL), Grenoble (France), and 2T at the Laboratoire Léon Brillouin (LLB), Saclay (France). The magnetic dispersion was measured on ARCS, with the sample's crystalline $c$-axis aligned along the incident beam, and with three incident neutron energies: $E_i = 70$ and 200 meV (at 5, 70, and 410 K) and $E_i = 50$ meV (at 10, 46, 72, 150, 250, 350, and 450 K). As in previous work on Hg1201 [27,28], the sample was only measured in one orientation, and the data were integrated along [0 0 1] due to the quasi-2D nature of the magnetic response in the lamellar cuprates. One side effect of this is that for any value of in-plane momentum transfer $\mathbf{q} = (H\ K)$, data with different energy transfer $\omega$ correspond to different values of out-of-plane momentum transfer $L$, as constrained by momentum and energy conservation. The dynamic magnetic susceptibility $\chi''(\mathbf{Q},\omega)$ was determined from the scattering cross-section by correcting for the anisotropic magnetic form factor [47,48] and the thermal Bose factor of the excitations, as described in previous work [27,28] and detailed in the Supplementary Information. Longitudinal polarization analysis was performed on IN20 [49] to extract the purely magnetic signal. Momentum scans across $\mathbf{q}_{AF}$ at select energies ranging from 9 to 30 meV, and energy scans at $\mathbf{q}_{AF}$ and two background momentum transfers were performed. The detailed temperature dependence of $\chi''(\mathbf{Q},\omega)$ at $\omega = 30$ meV (Fig. S9) and the elastic scattering (Fig. 5a) were measured on 2T.

The $q = 0$ magnetism measurements were carried on the D7 diffractometer at ILL [50], equipped with a cryofurnace, and with an incident neutron wavelength of 3.1 Å. The sample was aligned in the $(H\ 0\ L)$ scattering plane, and the alignment was checked using the OrientExpress diffractometer at ILL. The temperature dependence of the scattering at (1 0 0) was tracked between 100 K (above the SC transition to avoid the neutron beam depolarization) and 430 K using $XYZ$ longitudinal polarization analysis to search for a possible magnetic signal coincident with the nuclear (1 0 0) Bragg reflection, as detailed in the Supplementary Information.

### Density functional theory calculations

Phonon dispersions and eigenvectors were calculated using the density functional perturbation theory implementation in the *abinit* package [51]. Exchange-correlation was approximated with the PBEsol[52] version of the generalized-gradient approximation and norm-conserving pseudopotentials were used for the valence-core interaction [53]. The energy cutoff was 1200 eV and the *k*-point grid for ground state and perturbation calculations was 12x12x6. Dynamical matrices were calculated on a uniform 4x4x2 *q*-point grid, corresponding to a 4x4x2 supercell in real space. The real-space force-constants were ported to *phonopy* [54] format using *abipy* [55] and then used in the *euphonic* package [56] to calculate neutron intensities.

## Acknowledgements

The work at the University of Minnesota was funded by the U.S. Department of Energy through the University of Minnesota Center for Quantum Materials, under Grant No. DE-SC0016371. A portion of this research used resources at the Spallation Neutron Source, a Department of Energy Office of Science User Facility operated by Oak Ridge National Laboratory. Beam time was allocated to ARCS on proposals IPTS-8000, IPTS-12985, IPTS-14429, and IPTS-16717. Part of this research used resources at the Institut Laue-Langevin, via beam lines IN20 and D7. A portion of this research used resources at the Laboratoire Léon Brillouin, a facility funded by the French Atomic Energy Commission (CEA) and National Center for Scientific Research (CNRS), via beam line 2T. T.S. and D.R. acknowledge support by the U.S. Department of Energy, Office of Basic Energy Sciences, Office of Science, under Contract No. DE-SC0024117.

## Competing Interests

The authors declare no competing interests.

## Author Contributions

M.G. conceived the research. Z.W.A., Y.T., V.N., M.K.C., C.J.D., and G.Y. performed crystal growth, characterization, and co-alignment. Z.W.A., Y.T., M.K.C., Y.S., D.B., and L.M.-T. performed the neutron scattering experiments. D.L.A., A.D.C., L.M.-T., P.S., Y.S., and P.B. were local contacts for the neutron scattering experiments. Z.W.A., Y.T. M.K.C., and D.B. carried out the data analysis. T.S. and D.R. performed the DFT calculation. Z.W.A. and M.G. wrote the manuscript with input from all authors.

## Additional Information

**Correspondence** and requests for materials should be addressed to Z.W.A. and M.G.

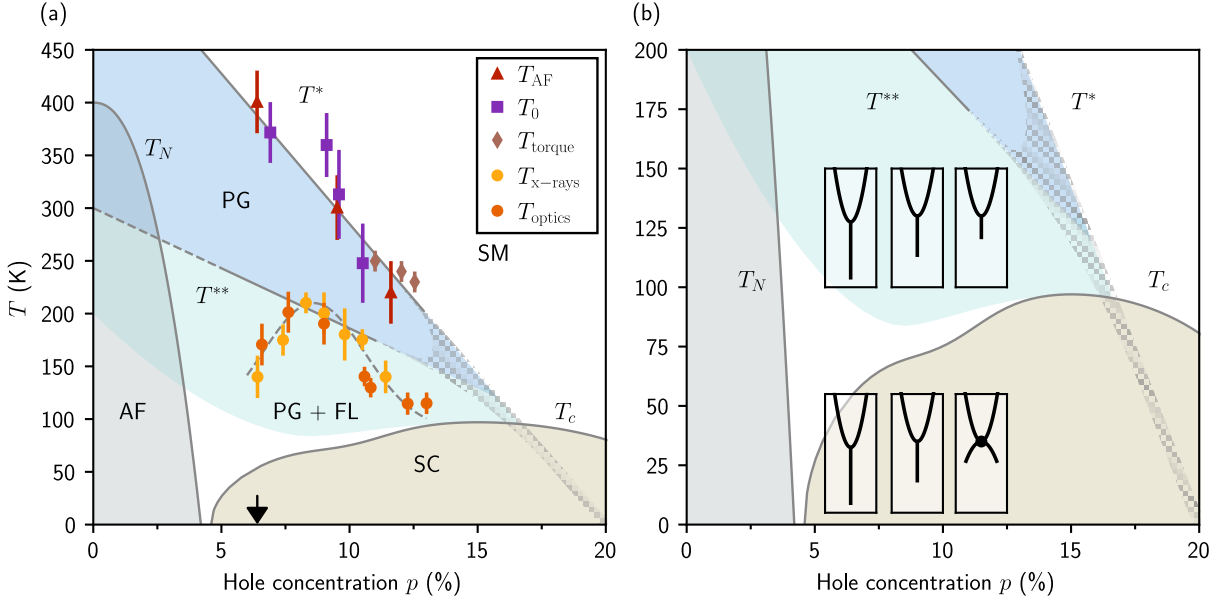

**Fig. 1 | Phase diagram**. (**a**) Phase diagram of Hg1201 ($p > 0.04$), extended to the undoped insulating state ($p = 0$) based on data for other cuprates [29], with antiferromagnetic (AF), superconducting (SC), pseudogap (PG) and Fermi-liquid (PG+FL) phases/regions. Lines indicate the characteristic pseudogap temperatures $T^*$ and $T^{**}$ and superconducting transition temperature $T_c$ obtained from charge transport measurements [39], and the approximate Néel temperature $T_N$. Red triangles: temperature $T_{AF}$ below which the response centered at $\mathbf{q}_{AF}$ is significantly enhanced (present work and Refs. [27,28]); purple squares; onset of $q = 0$ magnetic order [31-33]; brown diamonds: onset of magnetic order from torque magnetometry [57]; yellow and orange circles: characteristic temperatures of the onset of short-range charge-density-wave (CDW) order obtained from x-ray [29] and optical[58] measurements. At the doping level of the present study ($p \approx 0.064$, indicated by arrow), CDW correlations are weak. (**b**) Schematic diagrams of magnetic excitation spectra for Hg1201-UD55 (present work), Hg1201-UD71 [28] and Hg1201-UD88 [27] at two temperatures (above and below $T_c$), showing the characteristics of the magnetic dispersion as a function of energy (vertical) and momentum transfer centered at $\mathbf{q}_{AF}$ (horizontal). Circle indicates the presence of a resonance in Hg1201-UD88, near optimal doping.

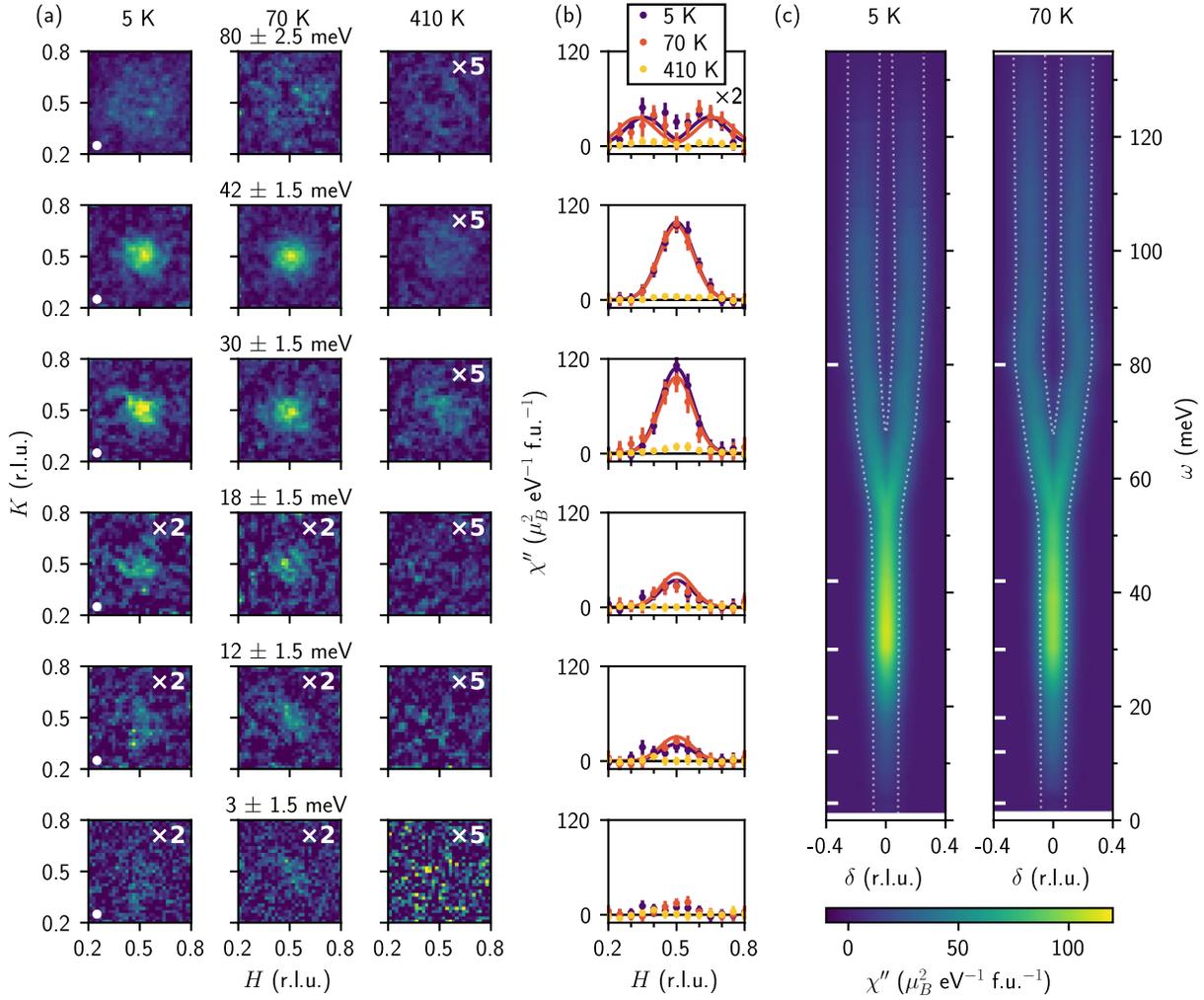

**Fig. 2 | Magnetic excitation spectrum featuring gapped commensurate and dispersive components.** (**a**) Constant-energy slices of magnetic susceptibility $\chi''(\mathbf{q})$ as a function of the 2D in-plane momentum transfer $\mathbf{q} = (H\ K)$ at $T = 5$ K (left), 70 K (center), and 410 K (right), in units of $\mu_B^2$ eV$^{-1}$ f.u.$^{-1}$ (same color scale as (c)). Data averaged over indicated energy ($\omega$) ranges. White text indicates that $\chi''$ has been multiplied by a factor to highlight more detail. White dots in lower-left corners indicate momentum resolution. Data with $\omega < 55$ meV ($\omega > 55$ meV) were collected with $E_i = 70$ meV ($E_i = 200$ meV). (**b**) Corresponding constant-$\omega$ cuts across $\mathbf{q}_{AF}$, averaged over {100} and {010} trajectories. The apparent susceptibility magnitude is slightly lower than the actual values due to the averaging over the binning range $q_\perp = [0.45, 0.55]$ r.l.u. (**c**) Magnetic dispersion extracted from 2D gaussian fits (see text), as a function of energy transfer ($\omega$) and incommensurability ($\delta = |\mathbf{q} - \mathbf{q}_{AF}|$) at $T = 5$ K (left) and 70 K (right). White dotted line indicates half-width-half-maximum of the response as a function of energy. White tick marks: $\omega$ values in (a, b). See Supplementary Information for data analysis details and additional data.

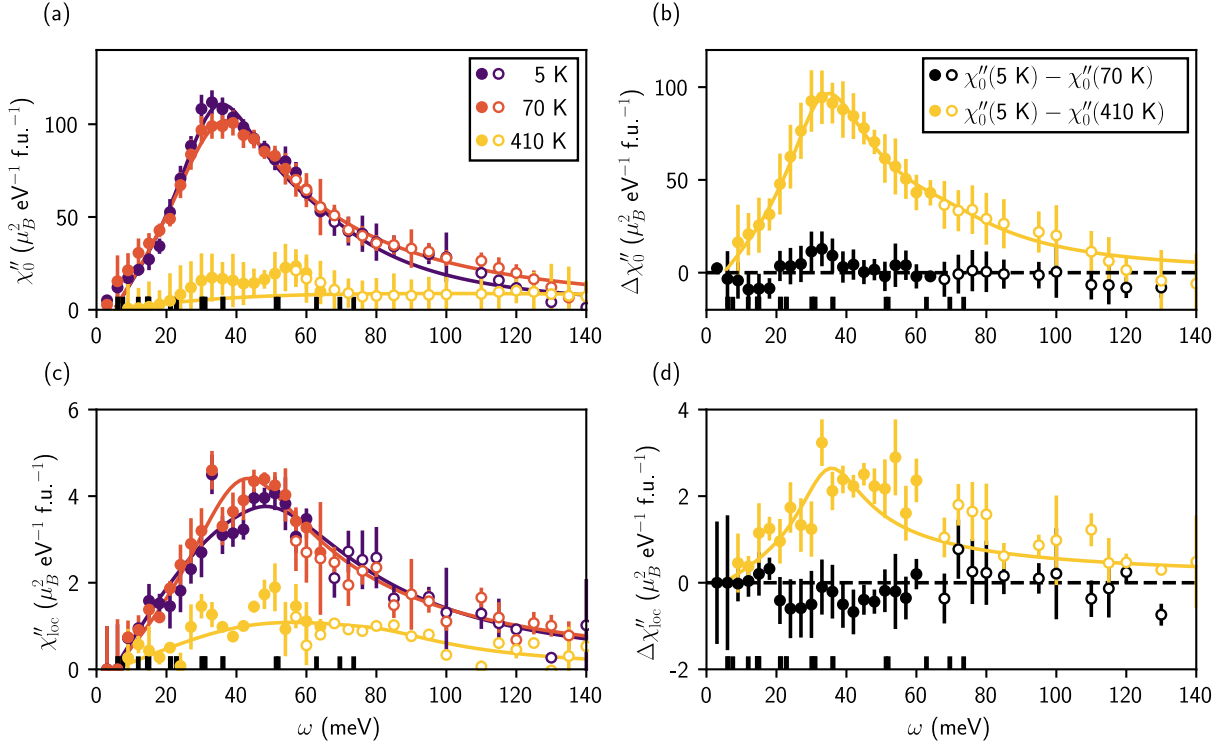

**Fig. 3 | Magnetic susceptibility amplitude and local susceptibility**. (**a**) Magnetic susceptibility amplitude $\chi''_0$, determined from 2D fits to background-subtracted data such as those in Fig. 2 (a). Filled circles: $E_i = 70$ meV. Open circles: $E_i = 200$ meV. (**b**) Difference between $\chi''_0$ at 5 K, deep in the superconducting state, and at 70 K (black) and 410 K (yellow). (**c**) Local (momentum-integrated) susceptibility $\chi''_{loc}$ (same colors and symbols as in (a)). (**d**) Difference in $\chi_{loc}$, using the same colors and symbols as (c). Black vertical bars in all panels: energies where DFT indicates phonon modes at $\mathbf{q}_{AF}$. At 410 K, the local maxima in $\chi''_0$ and $\chi''_{loc}$ at ~ 30 and ~ 50 meV correspond to phonon crossings and indicate the inability of our analysis to completely remove unwanted phonon contributions and/or to discern an enhanced magnetic response due to nontrivial electron-lattice coupling effects at these energies. The small enhancement in $\chi''_0$ near 30 meV across $T_c = 55$ K is consistent with the existence a small magnetic resonance, but also coincides with phonon crossings, and hence may be an electron-lattice-coupling effect or spurious. Solid lines: guides to the eye.

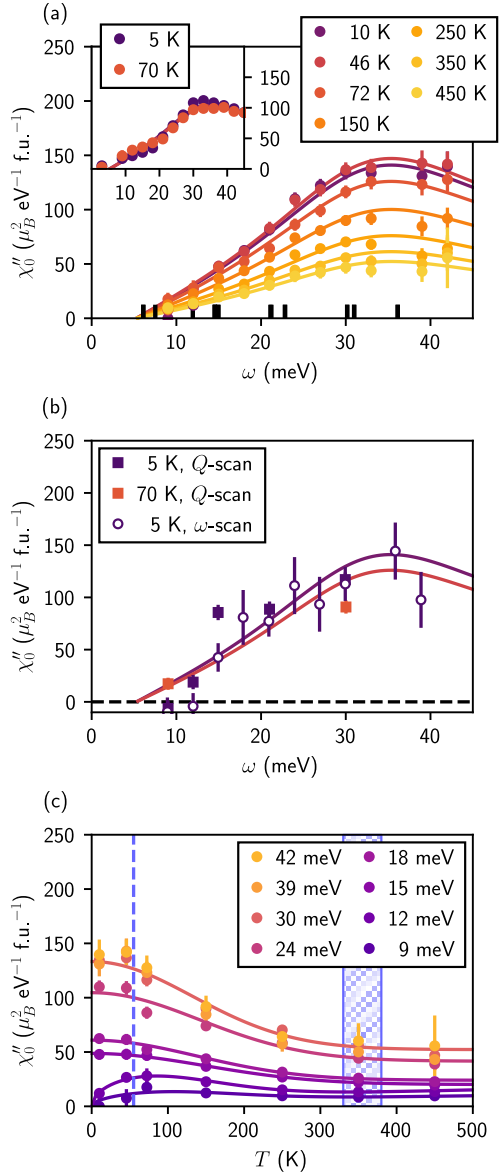

**Fig. 4 | Additional time-of-flight results and complementary triple-axis neutron data.** (**a**) Energy dependence of $\chi''_0$ measured with $E_i$ = 50 meV at seven temperatures. Solid lines: guides to the eye. Inset: $\chi''_0$ obtained with $E_i$ = 70 meV at 5 and 70 K (same data as in Fig. 3a). Black vertical bars: energies where DFT indicates phonon crossings at $\mathbf{q}_{AF}$. (**b**) Comparison of energy dependence of $\chi''_0$ (at $T$ = 10 K and 72 K; solid lines same as in (a)) with spin-polarized triple-axis neutron scattering results in the spin-flip channel at similar temperatures (squares: data obtained at 5 K and 70 K from momentum scans at select energies; circles: data obtained at 5 K from energy scans at $\mathbf{q}_{AF}$ and two background momentum transfers; see also Figs. S6 and S7 in Supplementary Information). The triple-axis data are scaled to match the time-of-flight result, which is obtained in absolute units. The combined data demonstrate the predominant magnetic nature of the signal, peaked at ~ 35 meV, and are fully consistent with the results in Fig. 3. (**c**) Temperature dependence of $\chi''_0$ at select energies integrated over ± 1.5 meV obtained with $E_i$ = 50 meV. Solid lines: guides to the eye. Vertical blue lines: $T_c$ = 55 K and $T^*$ ~ 370 K (see Fig. 1).

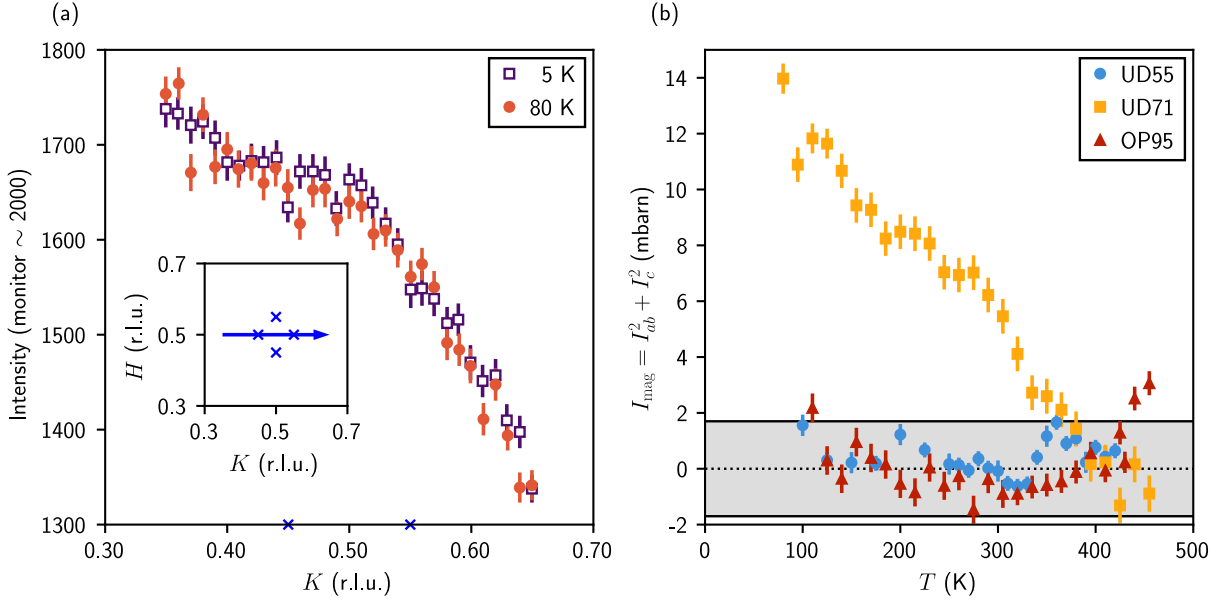

**Fig. 5 | Absence of static SDW/stripe and q = 0 order**. (**a**) Triple-axis momentum scan, across $\mathbf{q}_{AF}$ along [0 1 0], at $\omega = 0$ at 5 K (black squares) and 80 K (red circles). Blue crosses: locations where SDW/stripe order is observed at a similar doping level in LSCO ($p = 0.06$) [11]. Inset: schematic of the momentum-scan range (blue arrow) and characteristic wave vectors for LSCO ($p = 0.06$, blue crosses). See Supplementary Information for more details. (**b**) Spin-polarized neutron scattering results for the magnetic scattering intensity at the (1 0 0) reflection, determined from the inverse flipping ratio; see Fig. S8 in the Supplementary Information for the full inverse flipping ratio data. The highly underdoped (Hg1201-UD55, blue circles) and optimally-doped (Hg1201-OP95, yellow squares; data from Ref. [33]) samples show no detectable static magnetic order, in contrast to the moderately underdoped sample (Hg1201-UD71, red triangles; data from Ref. [33]). All data were obtained with the D7 spectrometer at ILL. The gray band corresponds to the 1.7 mbarn upper bound for the magnetic intensity of Hg1201-UD55 (see main text).

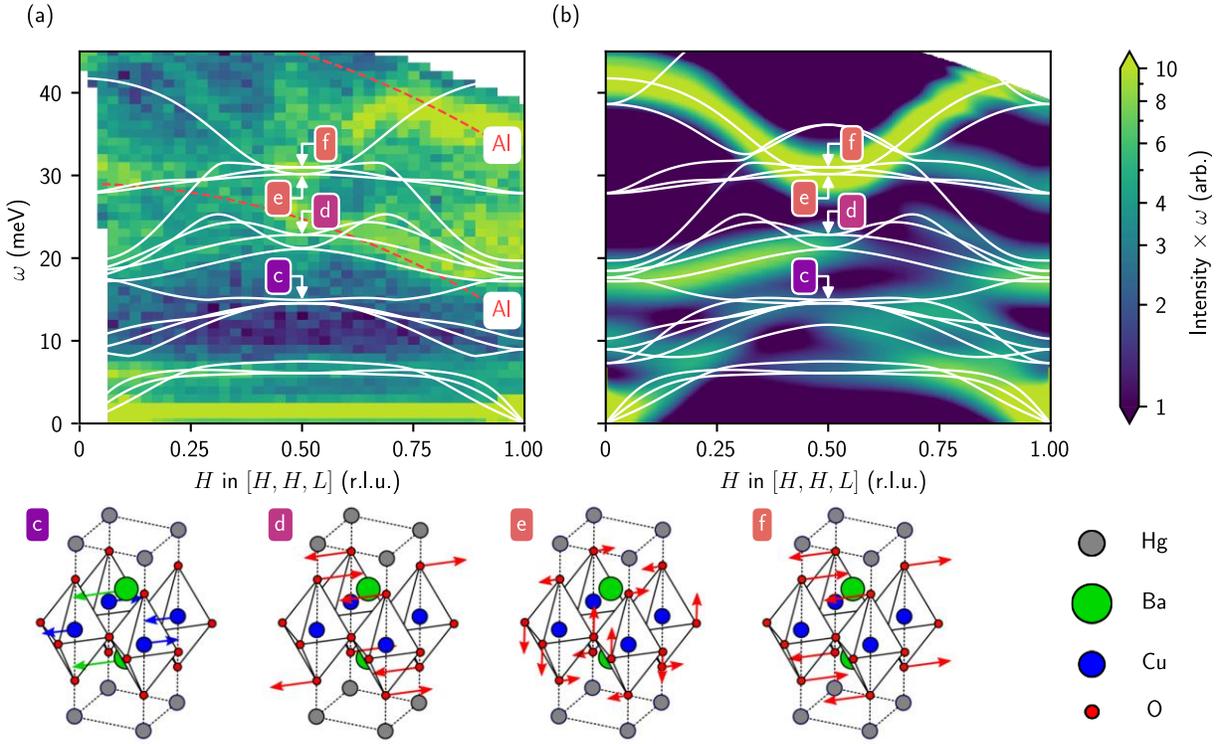

**Fig. 6 | Comparison of phonon modes with DFT results.** (**a**) Scattering intensity *without* background subtraction as a function of energy and momentum transfer along [1 1 0], integrated over $q_\perp = \pm 0.05$ (r.l.u.) along [1 -1 0] and over a wide range in $L$ (see Supplementary Information for details). The time-of-flight neutron data were collected at $T = 10$ K with $E_i = 50$ meV. Select calculated phonon modes are overlayed. Red dashed lines indicate spurious scattering from Aluminum in the sample holder. (**b**) Calculated phonon intensities in the same momentum range as the data in (a), with all phonon modes below 45 meV overlayed. (**c-f**) Select phonon eigenvectors at $\mathbf{Q} = (0.5\ 0.5\ L)$, with energies indicated in (a, b).

Supplementary Information for:

# Gapped commensurate antiferromagnetic response in a strongly underdoped model cuprate superconductor

Z. W. Anderson[1,*], Y. Tang[1], V. Nagarajan[1], M. K. Chan[1,†], C. J. Dorow[1], G. Yu[1], D. L. Abernathy[2], A. D. Christianson[3], L. Mangin-Thro[4], P. Steffens[4], T. Sterling[5], D. Reznik[5], D. Bounoua[6], Y. Sidis[6], P. Bourges[6], and M. Greven[1,*]

[1] *School of Physics and Astronomy, University of Minnesota, Minneapolis, MN 55455, USA*
[2] *Neutron Scattering Division, Oak Ridge National Laboratory, Oak Ridge, TN 37831, USA*
[3] *Materials Science and Technology Division, Oak Ridge National Laboratory, Oak Ridge, TN 37831, USA*
[4] *Institut Laue-Langevin, 71 avenue des martyrs, Grenoble 38000, France*
[5] *Department of Physics, University of Colorado Boulder, Boulder, Colorado 80309, USA*
[6] *Laboratoire Léon Brillouin, CEA-CNRS, Université Paris-Saclay CEA-Saclay, Gif sur Yvette 91191, France*
† *Present address: Pulsed Field Facility, National High Magnetic Field Laboratory, Los Alamos National Laboratory, Los Alamos, New Mexico 87545, USA*
* and01942@umn.edu, greven@umn.edu

**Content**

**Time-of-flight data analysis**

**Polarized neutron scattering**

**Investigation of q = 0 magnetism**

**Additional triple-axis scattering results**

**Additional phonon eigenvectors**

**Interdependence of ω, H, K, and L**

**References**

**Time-of-flight data analysis**

Time-of-flight (TOF) data were collected at the ARCS spectrometer at the Spallation Neutron Source, Oak Ridge National Laboratory. Raw data in the form of constant-energy slices in the vicinity of the 2D wave vector $\mathbf{q}_{AF}$ were analyzed after averaging the data in the four equivalent first magnetic Brillouin zones centered at $\mathbf{q}_{AF}$ = (0.5 0.5), (0.5 -0.5), (-0.5 0.5) and (-0.5 -0.5). Supplementary Figure 1 demonstrates the method used to estimate and subtract an energy (ω) and wave-vector (**Q**) dependent "background" $BG(\omega,\mathbf{Q})$ that includes phonons and powder rings. $BG(\omega,\mathbf{Q})$ is taken to be the average of the intensity at a given magnitude $Q$ at fixed energy transfer, with the data around $\mathbf{q}_{AF}$ excluded from the determination of the additional contributions (this method is referred to as Method 2 in Ref. S1). This background subtraction process very efficiently removes azimuthally symmetric spurious scattering, such as from phonon modes in the aluminum sample holder, but is less effective at removing some strongly-dispersing phonon modes of the sample.

The processed data were then converted to absolute units by comparison with a Vanadium standard. The imaginary part of the dynamic magnetic susceptibility, $\chi''(\mathbf{Q}, \omega)$, is related to the raw scattering cross section via

$$\frac{d^2\sigma}{d\Omega dE} = \frac{2(\gamma r_e)^2}{\pi g^2 \mu_B^2} \frac{k_f}{k_i} |F(\mathbf{Q})|^2 \frac{\chi''(Q,\omega)}{1 - \exp(-\omega/k_B T)} \quad (S1)$$

where $(\gamma\, r_e)^2$ = 0.2905 barn sr$^{-1}$, $k_f$ and $k_i$ are the final and incident neutron wave-vector magnitudes, respectively, and $|F(\mathbf{Q})|^2$ is the anisotropic Cu$^{2+}$ magnetic form factor[S2,S3]. The energy and momentum transfers are restricted by the energy and momentum conservation laws. Therefore, with fixed incident neutron energy, and for a given 2D momentum transfer $\mathbf{q} = (H\, K)$, the momentum transfer component $L$ perpendicular to the CuO$_2$ sheets depends on the energy transfer ω. Since the cuprates are quasi-2D materials, and the magnetic response in single-CuO$_2$-layer compounds is known to only weakly depend on $L$, we effectively consider the cross section to be independent of $L$. Neutron absorption was not considered in the unit conversion process.

The background-subtracted data were fit to a 2D gaussian function, as described in the main text and detailed in Fig. S2, in order to extract the peak susceptibility $\chi_0''$, incommensurability δ, and full-width-at-half-maximum (FWHM) $2\kappa$ of the magnetic response. Due to the relatively weak signal at 410 K, the $E_i$ = 70 and 200 meV data were fit with $2\kappa$ and δ fixed to their average 5 K and 70 K values to obtain the amplitude $\chi_0''$.

We analyzed data at various energy transfers: from ω = 9 meV to 54 meV in 3 meV steps, with a +/- 1.5 meV bin range; from ω = 60 meV to 75 meV in 5 meV step, with a +/- 2.5 meV bin range; from ω = 85 to 95 meV in 10 meV steps, with a +/- 5 meV bin range; from 115 to 130 meV in 15 meV steps, with a +/- 7.5 meV bin range. Figures 2, S3, S4, and S5 show extensive data at $T$ = 5 K, 70 K, and 410 K. Figures 2a and S3 demonstrate that no coherent AF signal is discernible at ω = 6 meV. As seen from these figures, the response remains commensurate up to ω ~ 54 meV, and then disperses outward and takes on a ring-like shape at higher energy transfers.

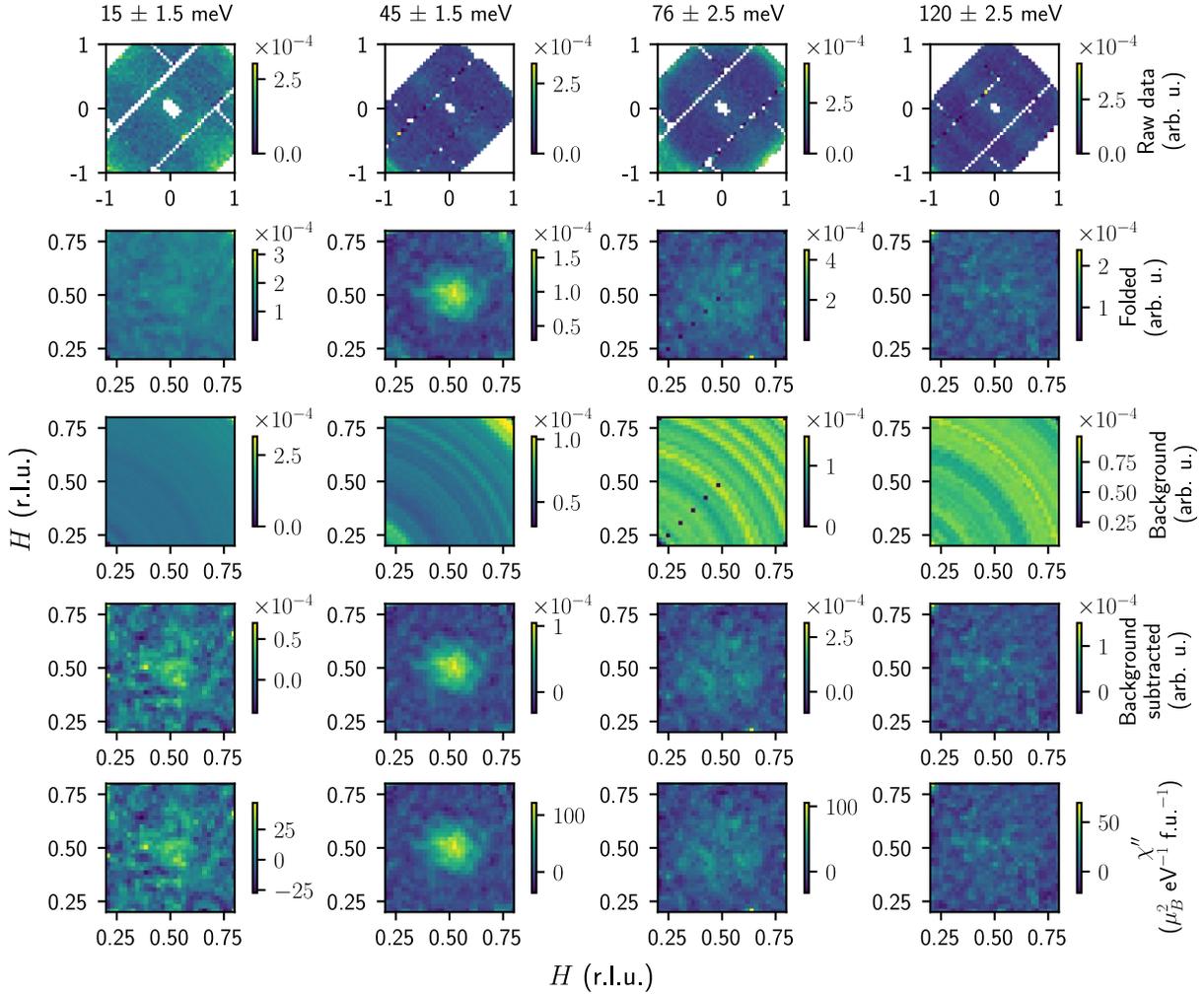

**Fig. S1 | Data processing procedure.** Example of the data processing and analysis steps at four values of ω (one per column), $T = 5$ K, with $E_i = 70$ meV (ω < 55 meV) and 200 meV (ω > 55 meV). The top row shows the raw data as a function of the 2D in-plane momentum transfer **q** = ($H\ K$), after integration over the out-of-plane momentum transfer $L$ and the energy binning range. The second row shows the data after folding into a single 2D reciprocal space quadrant (i.e., $-H \to H$, $-K \to K$). The third row shows the azimuthally symmetric background estimated as explained in the supplementary text. The fourth row shows the data after subtraction of this symmetric contribution. The bottom row shows the data after conversion to the imaginary part of the dynamic susceptibility.

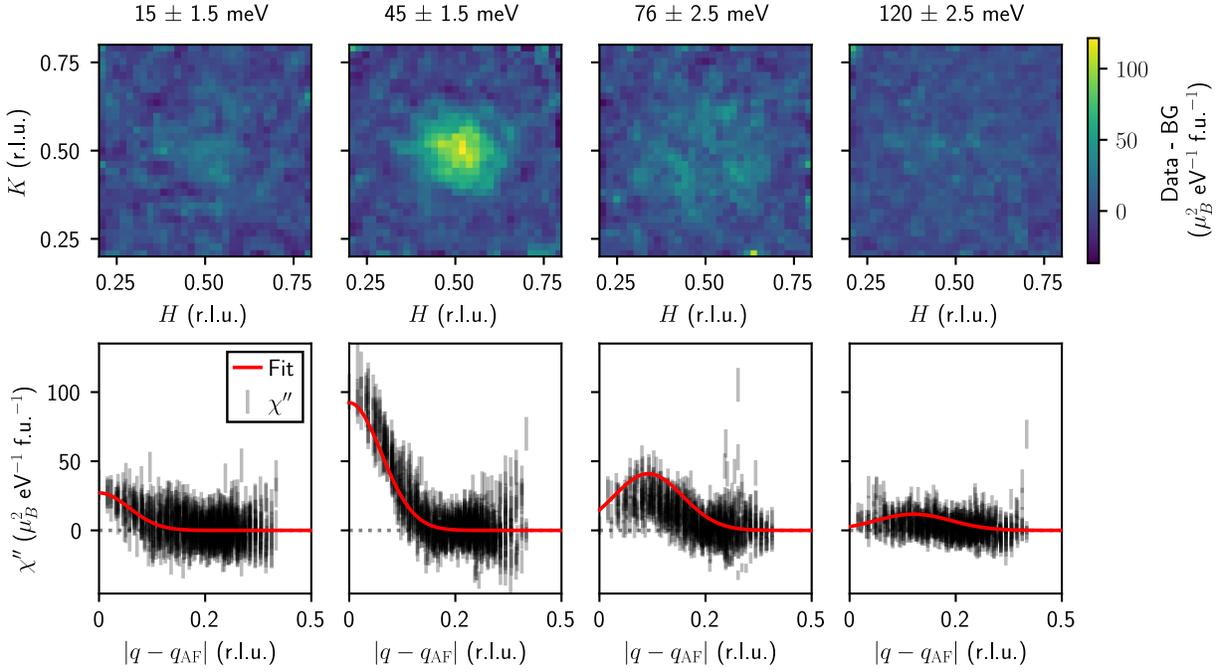

**Fig. S2 | Example gaussian annulus fit results.** Example of the results of fitting the gaussian annular function to the background-subtracted susceptibility data, as described in the supplementary text, at three values of ω and $T$ = 5 K. The top row shows the results for χ" after background subtraction (equivalent to the last row of Fig. S1). The bottom row shows the same χ" results as black vertical lines indicating the statistical uncertainty range for each pixel, plotted as a function of |$q$ - $q_{AF}$|, i.e., as a radial cross-section through the gaussian annulus, and the red line shows the fitted gaussian annulus.

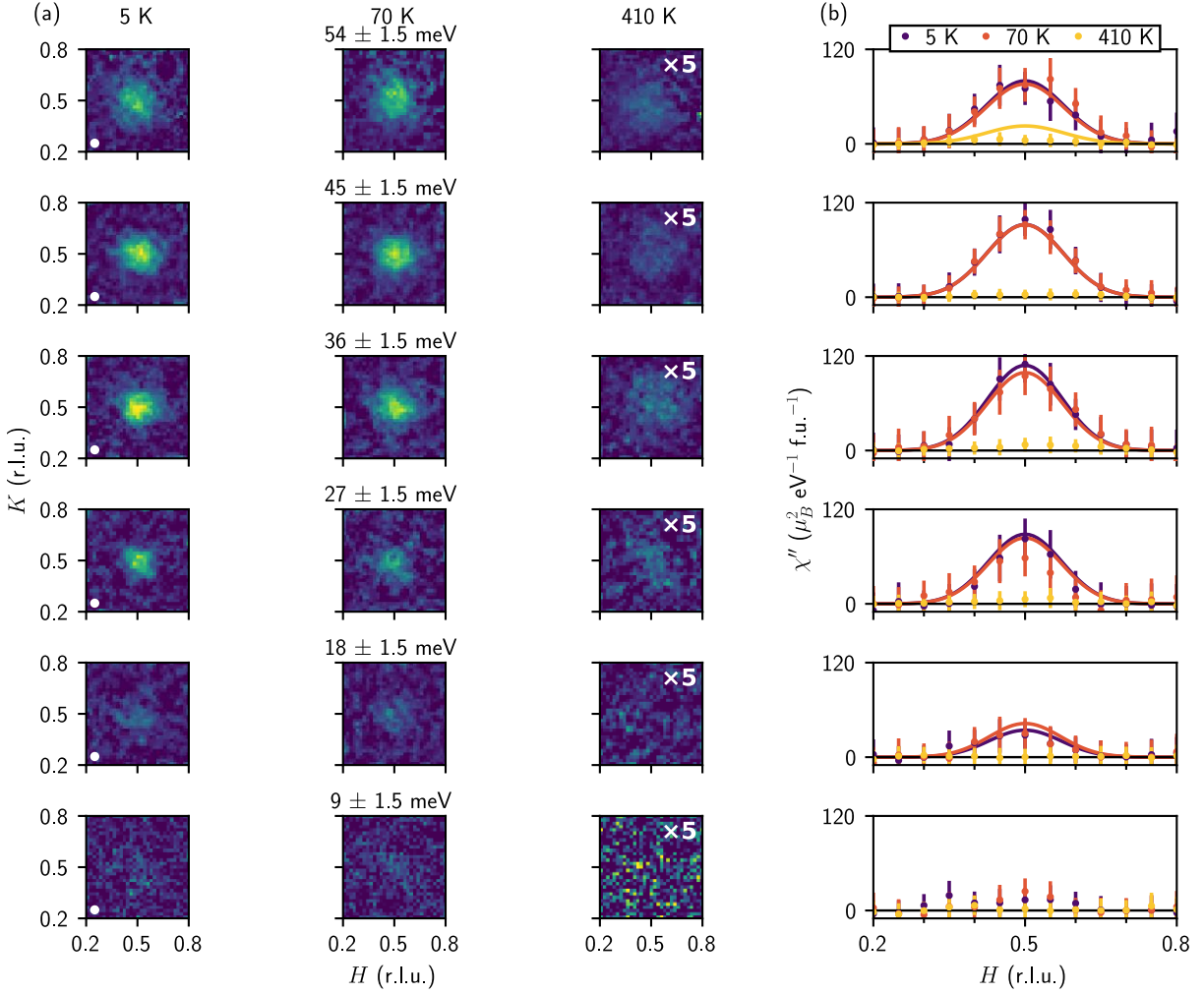

**Fig. S3 | Magnetic excitations - additional data at low energy transfers ($E_i$ = 70 meV).** (a) Constant-energy slices ($\omega$ < 55 meV) of magnetic susceptibility $\chi''(\mathbf{q})$ as a function of the 2D in-plane momentum transfer $\mathbf{q} = (H\,K)$ at $T$ = 5 K, 70 K and 410 K, in units of $\mu_B^2$ eV$^{-1}$ f.u.$^{-1}$ (same color scale as in Fig. 2c). Data averaged over indicated energy ranges. White text indicates where $\chi''$ has been multiplied by a factor to show more detail. White dots (left-most panels): momentum resolution. Data complement those in Fig. 2. (b) Corresponding constant-energy cuts across $\mathbf{q}_{AF}$ averaged over {100} and {010} trajectories across $\mathbf{q}_{AF}$ within a transverse momentum range from 0.46 to 0.54 r.l.u. Solid lines: Gaussian fits to full 2D slices of data. The intensity is slightly lower than the actual value due to averaging within the binning region.

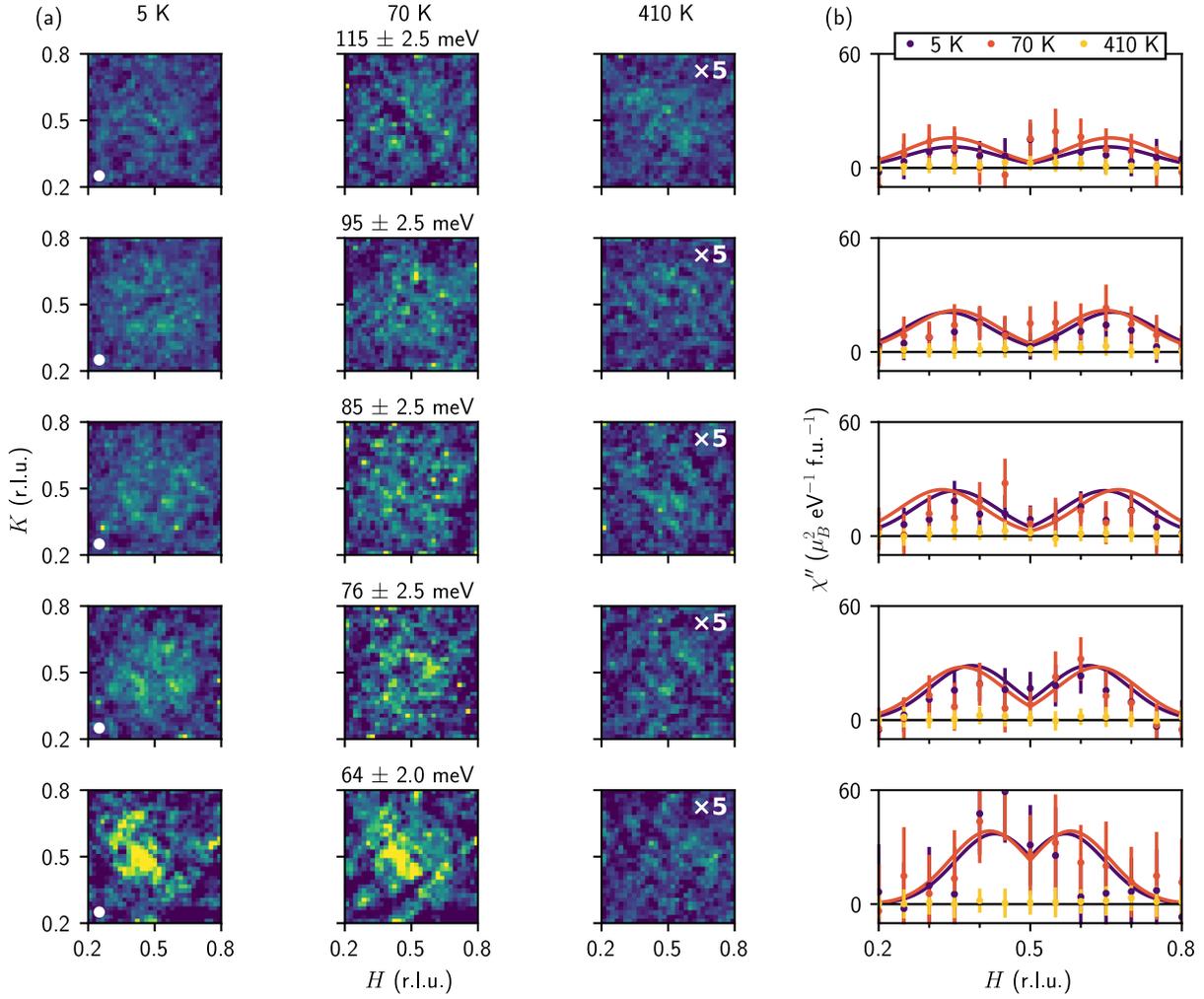

**Fig. S4 | Magnetic excitations – additional data at high energy transfers ($E_i$ = 200 meV). (a)** Constant-energy slices ($\omega$ > 55 meV) of magnetic susceptibility $\chi''(\mathbf{q})$ as a function of the 2D in-plane momentum transfer $\mathbf{q} = (H, K)$ at $T$ = 5 K, 70 K and 410 K, in units of $\mu_B^2$ eV$^{-1}$ f.u.$^{-1}$ (same color scale as in Fig. 2c). Data averaged over indicated energy ranges. White text indicates where where $\chi''$ has been multiplied by a factor to show more detail. White dots (left-most panels): momentum resolution. Data complement those in Fig. 2. **(b)** Corresponding constant-energy cuts averaged over {100} and {010} trajectories across $\mathbf{q}_{AF}$ within a transverse momentum transfer range from 0.46 to 0.54 r.l.u. Solid lines: Gaussian fits to full 2D slices of data. Due to the averaging over momentum transverse to {100} and {010} trajectories, the apparent signal in the right panels is slightly lower than the actual value shown in the contour plots.

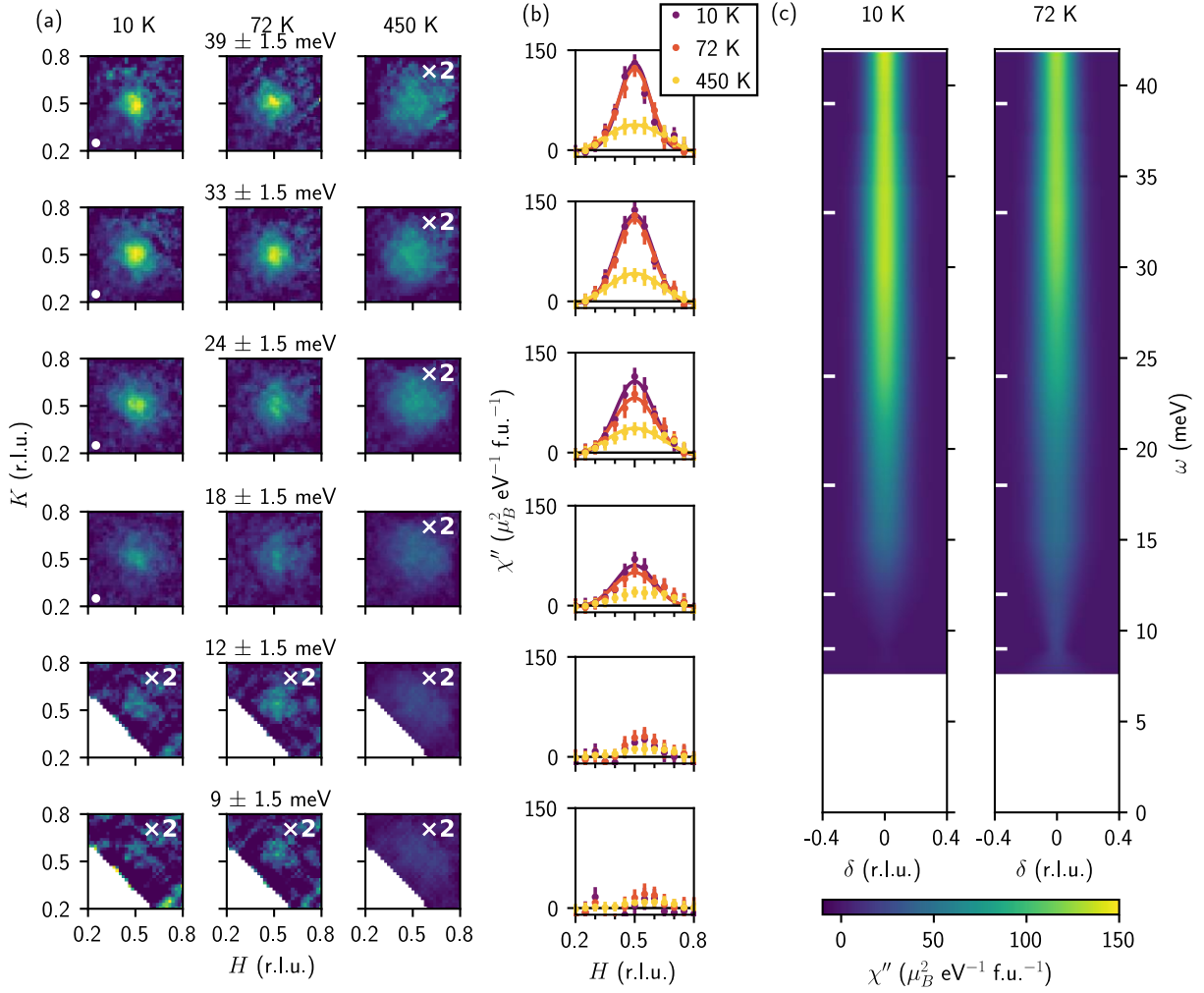

**Fig. S5 | Magnetic excitation spectrum with $E_i$ = 50 meV.** In contrast to Fig. 2 in the main text, this figure shows results obtained with $E_i$ = 50 meV. The corresponding findings for $\chi''_0$ are summarized in Fig. 4. (**a**) Constant-energy slices of magnetic susceptibility $\chi''(\mathbf{q})$ at $T$ = 5 K (left), 72 K (center), and 450 K (right), in units of $\mu_B^2$ eV$^{-1}$ f.u.$^{-1}$ (same color scale as (c)). Data are averaged over the indicated energy ranges. White text indicates where $\chi''$ has been multiplied by a factor to show more detail. White dots in lower-left corners indicate approximate instrument momentum resolution. (**b**) Corresponding constant-$\omega$ cuts across $\mathbf{q}_{AF}$, averaged over {100} and {010} trajectories. The apparent peak magnitude is slightly lower than the actual values due to the averaging over the orthogonal binning range from 0.45 to 0.55 r.l.u. (**c**) Magnetic response extracted from 2D gaussian fits to data such as those in (a) (see text) as a function of energy transfer $\omega$ and incommensurability $\delta = |q - q_{AF}|$, at $T$ = 5 K (left) and 72 K (right). White tick marks indicate the centers of the energy bins in (a, b).

**Polarized inelastic neutron scattering.**

Spin-polarized neutron scattering measurements of the magnetic excitation spectrum were carried out on the triple-axis spectrometer IN20 at the Institut Laue-Langevin, France. Heusler alloy crystals were used as a monochromator and analyzer to select the initial and final neutron energies and spin polarizations. The polarization of the neutron beam in the vicinity of the sample was maintained by CryoPAD[S4], which provides high stability and reproducibility of the neutron spin polarization. Figure S6 shows three-point momentum scans (at **q** = (0.3 0.3), (0.5 0.5) and (0.7, 0.7) r.l.u.) at various fixed energy transfers, whereas Fig. S7 shows energy scans at the same three momentum transfers.

We use longitudinal polarization analysis (LPA) to extract the magnetic contribution to the measured intensities. We define a Cartesian coordinate system based on the relative orientations of the neutron spin polarization (**P**) at the sample: **X** ∥ **Q**, where the **Q** is the scattering wave-vector; **Y**⊥**Q** and in the scattering plane; **Z** is the direction perpendicular to the scattering plane. In the absence of chiral magnetic correlations, the measured spin-flip (SF) scattering intensities in the three principal spin-polarization geometries are given by the following relations[S5]:

$$I_{P\|X}^{SF} = \frac{2}{3}N_{inc,spin} + M_{P\|Y} + M_{P\|Z} + BG$$
$$I_{P\|Y}^{SF} = \frac{2}{3}N_{inc,spin} + M_{P\|Z} + BG$$
$$I_{P\|Z}^{SF} = \frac{2}{3}N_{inc,spin} + M_{P\perp Q} + BG$$

where $N_{inc,spin}$ is the spin incoherent cross section, $M$ the magnetic cross section, and $BG$ the background contribution. We can calculate $M$ from the above equations using $M = M_{P\|Y} + M_{P\|Z} = I_{LPA} = 2 * I_{P\|X} - I_{P\|Y} - I_{P\|Z}$. The LPA in Fig. S6 clearly shows magnetic signal above $\omega = 9$ meV for $T = 5$ K, and possibly a small magnetic signal at $\omega = 9$ meV and $T = 70$ K. Figure S7a shows additional energy scans at the same three momentum transfers for incident neutron spin polarization along $X$. As summarized in Figs. 4b and S7b, the polarized-neutron data clearly demonstrate the predominant magnetic nature of the TOF signal. Moreover, whereas in the TOF phonons "contaminate" the local susceptibility at ~ 30 meV, the polarization analysis confirms the magnetic susceptibility peak around this energy.

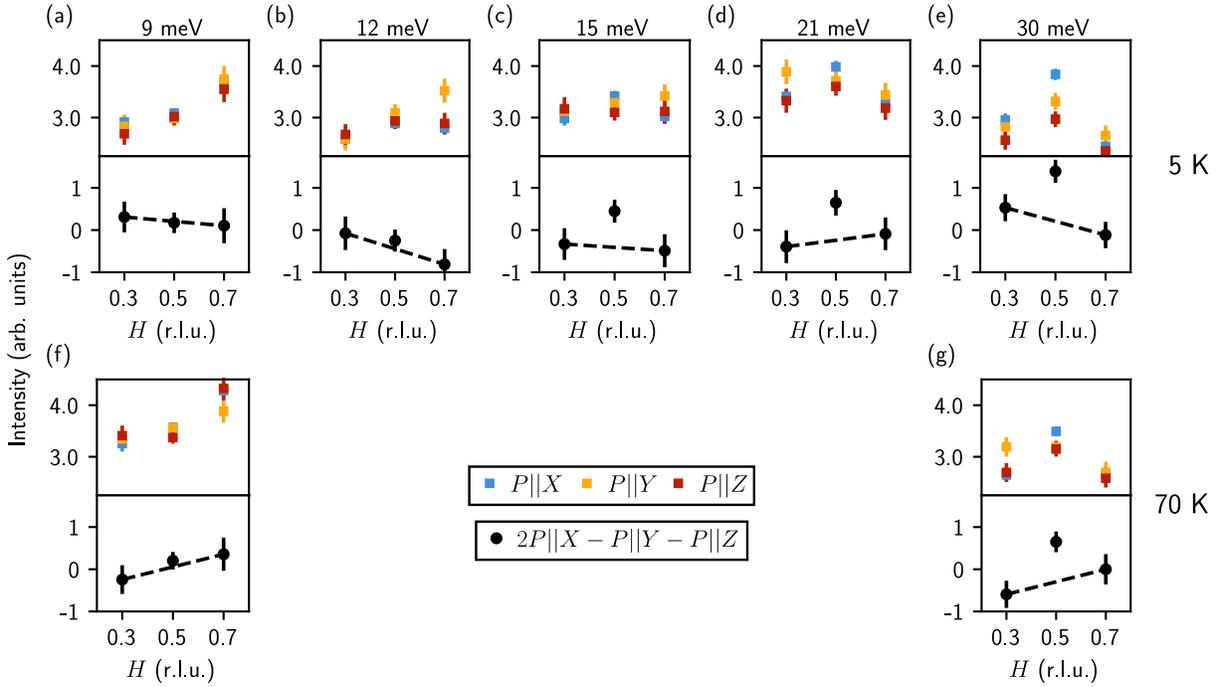

**Fig. S6 | Momentum scans at several energy transfers from the IN20 polarized neutron scattering measurement.** Top and bottom panels show measured intensity at 5 K and 70 K, respectively, at $\mathbf{q}_{AF} = (0.5, 0.5)$ r.l.u. and at the two background positions $\mathbf{q} = (0.3, 0.3)$ and $(0.7, 0.7)$ r.l.u. Blue, yellow, and red squares in the top row of each panel represent measurements with incident neutrons polarized along the $X$, $Y$ and $Z$ directions, where $X$ is parallel to the momentum transfer $\mathbf{Q}$, $Y$ within the scattering plane and perpendicular to $X$, and $Z$ perpendicular to both $X$ and $Y$. Black circles in the bottom row of each panel show the longitudinal polarization analysis (LPA) result, which is calculated by the formula $I_{LPA} = 2 * I_{P\|X} - I_{P\|Y} - I_{P\|Z}$. Dashed lines indicate linear interpolation of background. There is magnetic signal above $\omega = 9$ meV at $T = 5$ K, and possibly already at $\omega = 9$ meV at $T = 70$ K. This magnetic signal is shown in Fig. 4b.

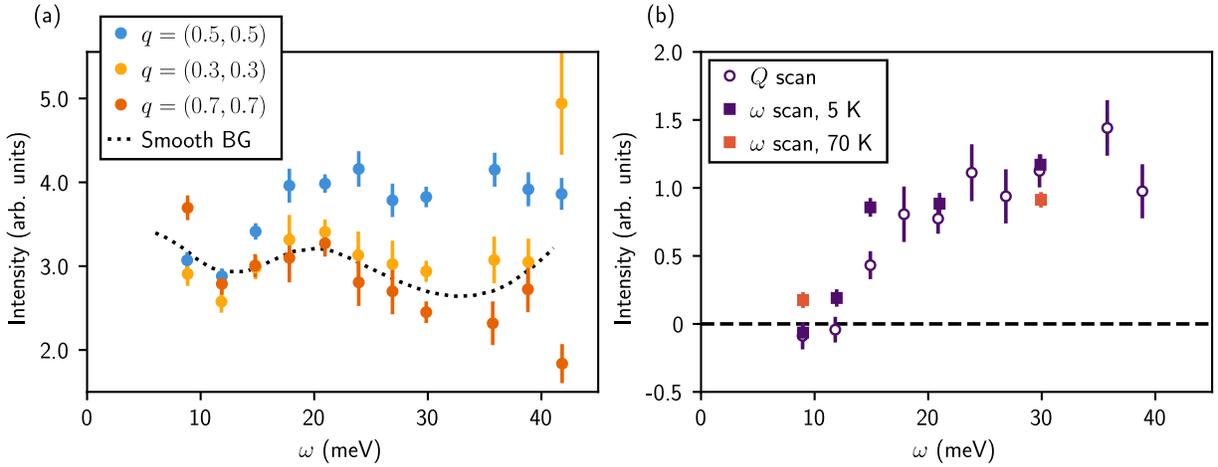

**Fig. S7 | Energy scans at three momentum transfers at $T$ = 5 K from the IN20 polarized neutron scattering measurement.** (**a**) Energy scan with incident neutron spin polarization along $X$ at three planar momentum transfers. As an estimate of the background level away from $\mathbf{q}_{AF}$ = (0.5, 0.5) r.l.u., the data at $\mathbf{q}$ = (0.3, 0.3) and (0.7, 0.7) r.l.u. are averaged and fit to a polynomial (dashed black line). (**b**) Energy dependence of magnetic intensity at $\mathbf{q}_{AF}$ obtained after removal of background estimate in (a) (open circles) and from momentum scans in Fig. S6 (purple and red filled squares, respectively). This magnetic signal is compared with the corresponding $E_i$ = 50 meV time-of-flight result in Fig. 4b.

**Investigation of q = 0 magnetism**

Measurements of the q = 0 magnetic signal were performed on the D7 cold-neutron diffractometer at the Institut Laue-Langevin (ILL), Grenoble, France. The initial polarization state of the neutron beam is prepared using a supermirror polarizer and a Mezei flipper that adiabatically flips the neutron spin. The neutron spin polarization (**P**) is maintained with a homogeneous guide field and controlled by a quadrupolar assembly at the sample position, while the final neutron spin polarization is analyzed using polarizing benders placed in front of a multidetector bank. Unlike the IN20 spectrometer, where **X** is set parallel to the scattering wavevector **Q**, D7 has fixed polarization directions for the neutron beam along **X**, **Y,** and **Z**. The angle between **Q** and **X** is $\alpha = \frac{\pi}{2} - \theta + \gamma$, where $\theta$ is the Bragg scattering angle and $\gamma$ is the angle between incident beam and **X**, which is determined to be approximately 43º on D7. Therefore, for an incoming beam wavelength $\lambda = 3.1 \text{Å}$, $\alpha = 108.1° \pm 5°$ for **Q** = (1 0 0), where the uncertainty comes from both the sample mosaic ($\theta$) and the instrument ($\gamma$).

Detailed measurements of the temperature dependence at the (1 0 0) Bragg peak were performed along the three neutron polarization directions (**X**, **Y**, and **Z**), in the spin-flip (SF) and the non-spin-flip (NSF) channel. Due to the weak amplitude of the expected magnetic intensity and imperfect polarization leading to a leakage from the NSF channel (where nuclear scattering dominates) to the SF channel, we followed the temperature dependence of the flipping ratio (FR):

$$FR = \frac{NSF_{X,Y,Z}}{SF_{X,Y,Z}}$$

The temperature dependence was measured after choosing the best bender-detector pair that matches the **Q**-range of interest in order to achieve the best possible flipping ratio of about 50 at 80 K. Figure S8 shows the prior result for temperature dependence of the inverse flipping ratio (1/FR) at (1 0 0) for Hg1201-UD71 and Hg1201-OP95[S6], along with new data for Hg1201-UD55, in the three (**X,Y,Z**) polarization channels. While the inverse flipping ratio exhibits an upturn below $T_0 \sim 370$ K for Hg1201-UD71, the corresponding data at higher and lower doping show a weak temperature dependence below 400 K, indicative of the absence of magnetic scattering within the detection limit of D7.

We analyzed the new data for Hg1201-UD55 following the same protocol as described in detail in the supplementary information in Ref. S6. In particular, the determination of the temperature dependence of the magnetic scattering at (1 0 0) shown in Fig. 5b was achieved using a "blind test" method. This method allows for a $q = 0$ magnetic signal with an order-parameter-like temperature dependence in a simultaneous fit of the data for all three samples and only assumes that all samples exhibit the same SF background temperature dependence.

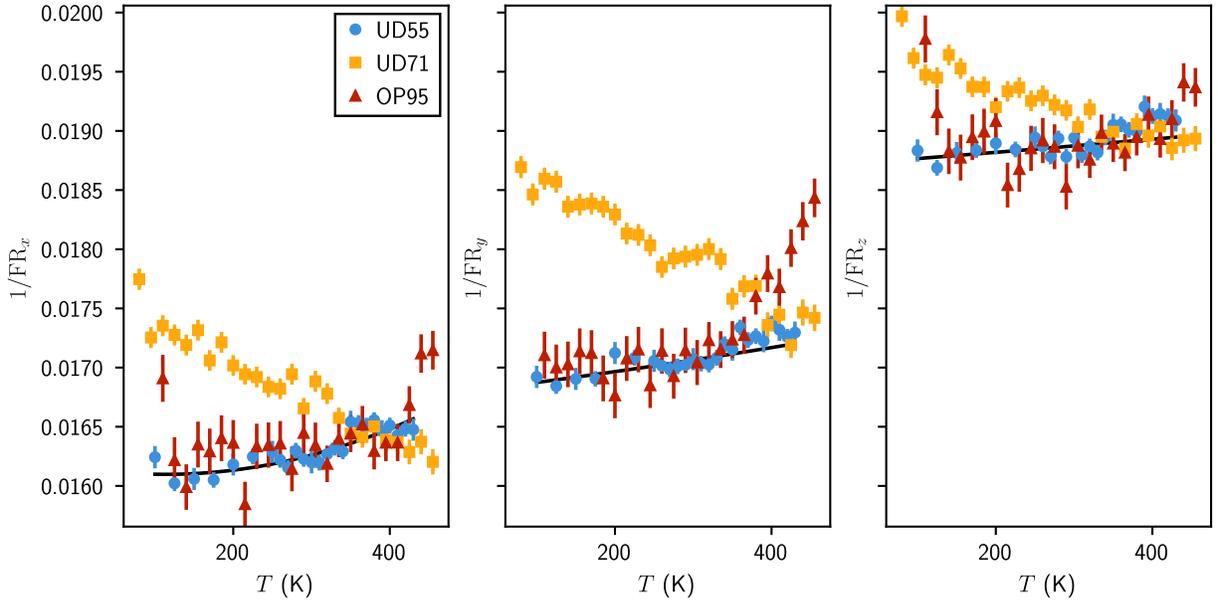

**Fig. S8 | Comparison of inverse flipping ratio from spin-polarized measurements of three Hg1201 samples.** Left: Inverse flipping ratio (1/FR) at the (1 0 0) reflection for the three polarization directions for Hg1201-UD55 (blue), Hg1201-UD71 (yellow), and Hg1201-OP95 (red). The latter two samples were measured previously[S6]. A magnetic signal is seen in the Hg1201-UD71 data from the upturn below $T_0 \sim 370$ K, but not for the other two samples. Solid black lines are smooth polynomial fits to the new Hg1201-OP95 and Hg1201-UD55 data, with less weight given to the high-temperature results, where the uncertainty in FR increases due to thermal effects on the sample.

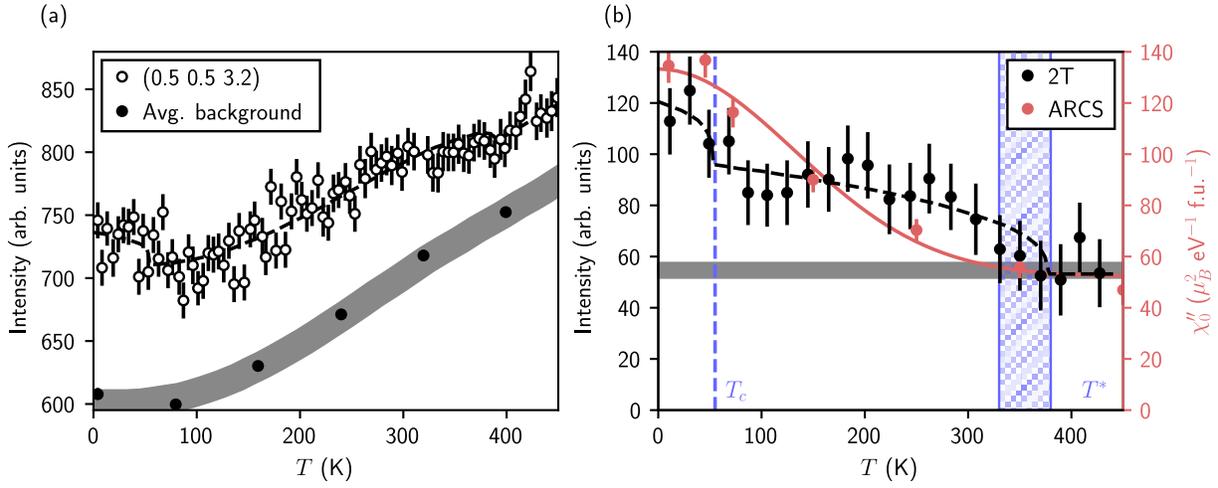

**Fig. S9 | Temperature dependence of the intensity at ω = 30 meV, obtained with the 2T triple-axis spectrometer at LLB.** (**a**) Open circles: intensity at **Q** = (0.5 0.5 3.2). Closed circles: background intensity estimated at six temperatures by averaging rocking scans within the range $H = 0.3 \pm 0.1$ r.l.u.. Grey shaded band is the resultant estimated background temperature dependence; the width of the band represents the estimated uncertainty. (**b**) Black circles: Temperature dependence of the magnetic intensity, extracted by subtracting estimated background from the intensities at **Q** = (0.5 0.5 3.2), as shown in (a). Horizontal grey band: Uncertainty in signal baseline due to the uncertainty in background. Black dashed line: guide to the eye. Red circles and line: Corresponding $E_i$ = 50 meV time-of-flight neutron data and guide to the eye from Fig. 4c. The combined data do not allow us to distinguish between a small resonance-like enhancement of the response below $T_c$ and a gradual intensity increase with decreasing temperature. Vertical blue lines/bands indicate estimates of $T_c$ and $T^*$ (same as in Fig. 4c).

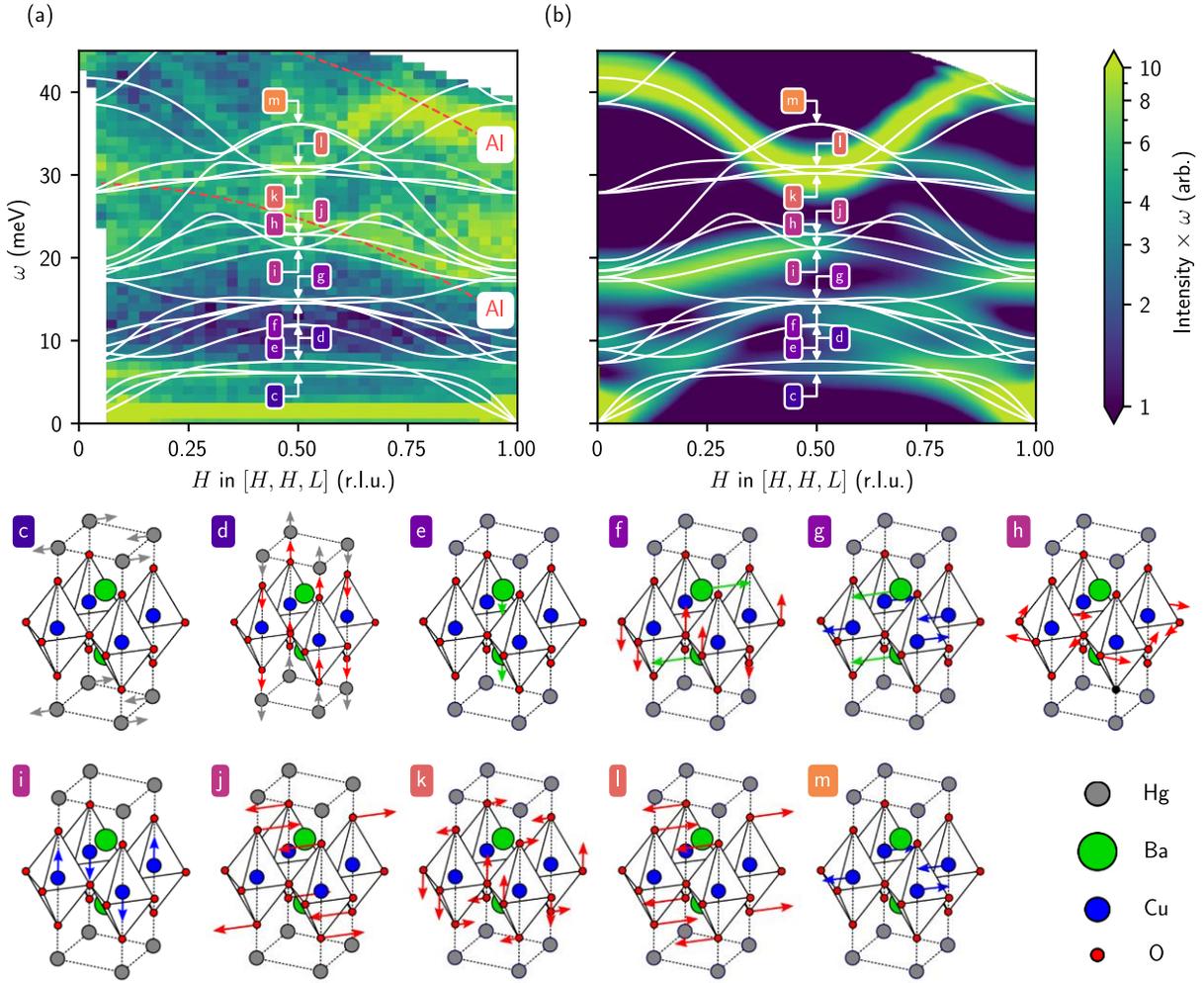

**Fig. S10 | Further comparison of phonon scattering and DFT calculation.** (**a**) Scattering intensity as a function of **q** along [1 1 0] and energy transfer $\omega$. (**b**) Phonon intensities from DFT in the same momentum range as the data in (a), with phonon modes overlayed. (**c-m**) Phonon eigenvectors at **Q** = (0.5 0.5 $L$) and select energies indicated in (a, b). In contrast to Fig. 6, all phonon modes that cross **Q** = (0.5 0.5 $L$) below $\omega$ = 45 meV are displayed. See main text for more detail.

**Interdependence of ω, H, K, and L**

During the neutron scattering measurements at ARCS, the sample was kept in a fixed orientation relative to both the incident beam and the detector, and data were collected in a three-dimensional volume parameterized by a two-dimensional scattering angle and a time-of-flight. Upon transformation to reciprocal space and energy transfer coordinates, the measured volume corresponds to a three-dimensional subset of the four-dimensional momentum (*H K L*) and energy transfer (ω) space, such that for given values of *H* and *K*, the neutron scattering cross section was only measured at a single value of *L* and a single value of *ω*, where *L* and *ω* are coupled according to conservation of energy and momentum, and the shape of the ARCS detector. In our data analysis, we assume that the only *L*-dependence of the magnetic scattering is due to the overall **Q**-dependence of the magnetic form factor (see Eq. S1). After converting the data to units of magnetic susceptibility, we integrate over *L*, which corresponds to a projection of the data onto a three-dimensional basis parameterized by *H*, *K*, and *ω*. Note that this means that planar cuts in *H-K-ω* space, e.g., those shown in Figs. 2a and S5a, correspond to measurements in which the value of *L* depends on *H*, *K*, and *ω*. This non-trivial dependence is taken into account when comparing the measurements to the DFT-derived phonon spectrum.